**Solvent-antisolvent interactions in metal halide perovskites**

Jose Roberto Bautista-Quijano,[ab] Oscar Telschow, [ab] Fabian Paulus,[bc] and Yana Vaynzof *[ab]

a.   Chair for Emerging Electronic Technologies, Technical University Dresden, Nöthnitzer Str. 61, 01187 Dresden, Germany
b.   Leibniz-Institute for Solid State and Materials Research Dresden, Helmholtzstraße 20, 01069 Dresden, Germany
c.   Center for Advancing Electronics Dresden, Technical University of Dresden, Helmholtz Str. 18, 01069, Dresden, Germany

*Email: yana.vaynzof@tu-dresden.de

The fabrication of metal halide perovskite films using the solvent-engineering method is increasingly common. In this method, the crystallisation of the perovskite layer is triggered by the application of an antisolvent during the spin-coating of a perovskite precursor solution. Herein, we introduce the current state of understanding of the processes involved in the crystallisation of perovskite layers formed by solvent engineering, focusing in particular on the role of antisolvent properties and solvent-antisolvent interactions. By considering the impact of the Hansen solubility parameters, we propose guidelines for selecting the appropriate antisolvent and outline open questions and future research directions for the fabrication of perovskite films by this method.

**Introduction**

Metal halide perovskites (MHPs) have attracted significant attention from the scientific community due to their remarkable optoelectronic properties such as high absorption coefficients,[1] large electron-hole diffusion lengths,[2] high charge-carrier mobilities[3] and low degree of energetic disorder.[4] These properties make MHPs particularly suited for application in electronics and optoelectronics with a broad range of devices already demonstrated.[5] These devices include solar cells,[6] light-emitting diodes,[7] photodetectors,[8] lasers,[9] field-effect transistors,[10] memristors[11] and scintillators.[12] Remarkably, the high optoelectronic quality of perovskites can be achieved despite the fact that they are commonly processed from solution in a simple deposition method and annealed at low temperatures, typically around 100 °C.[13]
It is noteworthy that while many solvent-free deposition methods have been utilised for MHPs fabrication such as thermal evaporation,[14–16] sputtering,[17,18] chemical vapour deposition[19,20] or other techniques,[21] the vast majority of the research community is focused on solution processing due to its simplicity.[22] Solution processing enables a large degree of control over the microstructure, crystalline structure and phase and orientation of perovskite thin films. For example, it has been observed that solution processing leads to more favourable microstructure (e.g., with increased average grain size) for a variety of perovskite compositions as compared to other deposition methods,[22] which is advantageous to the device performance. Importantly, solution-processing enables the introduction of additives, which have been shown to play a crucial role in enhancing both the efficiency and stability of perovskite devices.[23,24]

The combination of these factors led to the observation that the device performance of solution processed devices typically surpasses that of those fabricated by other means, which further motivates the use of solution processing for future research. The fabrication of MHP thin films by solution processing in a laboratory setting typically involves spin-coating of a perovskite precursor solution through one of the following three approaches: the so-called one-step, two-step, and the solvent-engineering method. The one-step method is the simplest and involves the mixing of all perovskite precursors in a single solution, which is then deposited onto the substrate in a single spin-coating step.[25,26] Such an approach is commonly applied to single cation compositions such as methylammonium lead triiodide ($MAPbI_3$),[27–30] methylammonium lead bromide ($MAPbBr_3$)[31] or caesium lead triiodide ($CsPbI_3$).[32–34] The two-step approach, which relies on the deposition of one of the precursors (typically a metal halide) first, followed by the deposition of a second precursor solution (typically organic halides) was among the earliest methodologies developed in the field.[35,36] Recently, this method attracted renewed attention due to its successful application in the fabrication of mixed cation-based perovskites.[37,38] Despite the relative success of these two methods, the most common method for perovskite layer fabrication is the solvent-engineering approach.[39] This method is based on the one-step approach, since all of the perovskite precursors are still deposited in a single spin-coating step, but with the application of an antisolvent shortly before the spin-coating is completed. The application of an antisolvent – a solvent in which the metal halide perovskite precursors are typically less soluble – results in a fast and uniform nucleation and solidification of the film.[40]

The ubiquitous use of the solvent-engineering method led to extensive investigations regarding the detailed processing parameters of the antisolvent application procedure, including the choice of solvent, its volume, its timing, temperature and duration.[41–46] In this feature article, we introduce the current state of understanding of the crystallisation process of the perovskite layer made by the solvent-engineering method, and focus on discussing the role of antisolvent properties and solvent-antisolvent interactions in determining the final structure and microstructure of the perovskite layer. Importantly, we introduce the importance of Hansen Solubility Parameters (HSP)[47] that should be taken into account when considering the interactions of the antisolvent with the perovskite host solvents. Finally, we provide an outlook and future directions for the further development of the solvent-engineering methods for fabrication of MHP layers.

**Solvent-Engineering in Metal Halide Perovskites**

The one-step method in the fabrication of MHPs relies on spontaneous nucleation and solidification of the precursor materials during the spin-coating, drying or annealing process. Since this spontaneous process is largely uncontrollable, the resulting films are typically characterized by large-scale nonuniformities in microstructure and quality. Dominant pinholes, laterally varying grain sizes or an incomplete precursor conversion are commonly observed for the one-step method.[48] These structural imperfections can in turn limit the overall device performance, increase non-radiative recombination or lower the stability of the perovskite. Only by precisely controlling the fabrication parameters such as atmosphere, humidity, temperature, drying rates and times, high quality films can be achieved. To tackle this challenge, in 2014 Xiao et al. introduced the addition of an antisolvent during the spinning process as an alternative fabrication route.[49] Initially referred to as "fast crystallization deposition", this procedure is depicted in Figure 1 and is now called the "solvent-engineering" method.

The term "antisolvent" is used for a solvent that provides significantly lower solubility for the metal and organic halide precursors and which is applied onto the wet substrate towards the end of the deposition process. By extracting the host solvents used to dissolve the precursor materials and lowering the precursor solubility, local supersaturation is achieved quickly, initiating the nucleation, precipitation and solidification of the dissolved materials into a film. Due to this very rapid process, the uniformity of the film is typically improved resulting in a better morphology and electronic quality, and subsequently, an improved performance and stability.[50–52] Host solvents - used to dissolve the precursors of MHPs are typically dimethylformamide (DMF), dimethyl sulfoxide (DMSO), often also in combination as solvent mixture, especially for MHPs containing cesium, since the common precursor CsI is insoluble in DMF. Further, also acetonitrile (ACN), N-methyl-2-pyrrolidone (NMP), gamma-butyrolactone (GBL), tetrahydrofuran (THF) and dimethylacetamide (DMAc) are among the most commonly investigated solvents[53–55]. Motivated by the search for a "green" solvent, even water has been successfully tested on certain perovskite compositions and device structures.[56]

The range of antisolvents employed by the perovskite research community is significantly broader, including polar, protic, aprotic and non-polar solvents with high and low dipole moments and boiling points. The most commonly employed solvents and antisolvents and their most common properties are listed in Table 1. We note that while most commonly being applied by pipetting during spin-coating,[57] antisolvents have also been reported to be applied in alternative methods such as dipping[43] and gas-assisted[58] or carrier-gas free spraying.[59]

The transition from a precursor solution to a perovskite film has been shown to normally proceed not directly into the perovskite phase, but often through the formation of intermediates.[60] These can include non-photoactive perovskite phases like ☐FAPbI$_3$,[61] iodo plumbates like [Pb$_3$I$_8$],[62] but also solvent containing intermediate phases. Especially Lewis-Acid-Lewis-Base-complexes as intermediates, like PbI$_2$-2·DMSO,[61] MAI-DMSO-PbI$_2$,[62] MAI-DMF-PbI$_2$,[63] PbI$_2$-2·DMF,[61,64] PbI$_2$-FAI-DMSO,[65] typically formed with the high-boiling point host solvents, often determine the final crystal grain size, microstructure and morphology of the perovskite film upon thermal annealing in the final step of the fabrication process.[66] While in the one-step method these intermediates can be manipulated by adjusting the surrounding atmosphere, the drying times and drying conditions, in the solvent-engineering method the type of antisolvent mainly determines the amount and composition of intermediates phases and consequently the perovskite film quality. For example, we recently demonstrated that the occurrence of a templating intermediate can influence grain orientation and consequently performance and stability but varies strongly based on the antisolvent used.[67]

The investigation of antisolvents has mainly been driven by the desire to improve device performance and stability, lower resource usage and reduce the environmental impact.[68,69] While several papers have compared antisolvents mainly based on the performance of the resulting perovskite solar cells, several attempts have been made to develop clear metrics to classify antisolvents suitability for given perovskite compositions and host solvents. The investigated parameters include among others boiling point, miscibility,[41] dielectric constant or relative permittivity,[50] dipole moment[70,71] and solubility of different precursors.[57,72] Over the past years an understanding has evolved that the molecular interactions between antisolvent, host solvent(s) and precursor material determine the overall solidification process and resulting perovskite microstructure and morphology. Despite the antisolvent-host solvent interaction often discussed in terms of one-dimensional parameters like polarity or protic/aprotic-character, the explanatory and predictive power for parameters like solubility and molecular interaction is limited, as they are an incomplete characterization of the molecular interactions,[73] dictating the host solvent–antisolvent–precursor interaction. As an example, we point out that despite similar dipole moments, the use of ethanol and chlorobenzene as antisolvents can lead to very different results from non-working, incomplete precursor conversion to high performance solar cells with PCE ~20%. Similarly, the use of ethanol and *tert*-butanol (*t*-BuOH) with similar boiling point results in starkly different perovskite film quality.[72,74]

A more sophisticated model to describe these interactions is given by the Hansen Solubility Parameters. Furthermore, these parameters are easily applicable to solvents, antisolvents and solvent- or antisolvent mixtures,[54] as well as precursors[75,76] and can be plotted in Hansen parameter space, giving a good visual and numerical guideline of the interactions between the component materials included in the perovskite formation. While some studies on perovskite solar cells have considered HSPs, these so far focused solely on solvent-perovskite interactions and thus, an overview of the positions in HSP space of commonly used solvents and antisolvents combinations which would be useful for the field has yet to be presented.

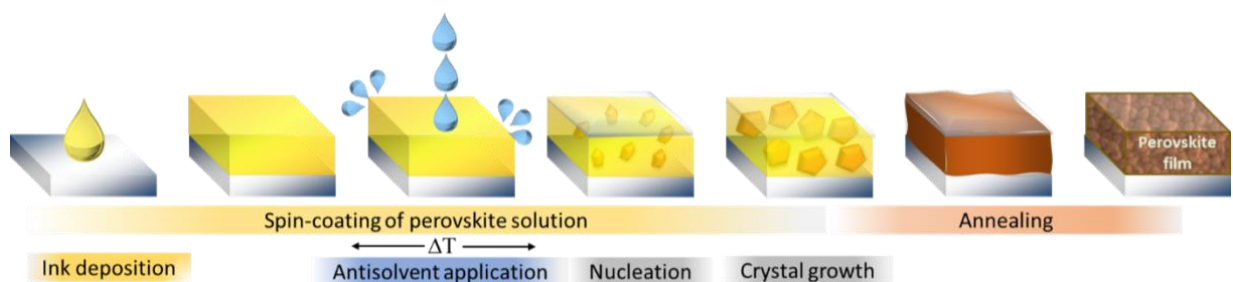

**Figure 1:** Schematic illustration of the fabrication of metal halide perovskite layers by the solvent-engineering method.

**Table 1:** Properties of common solvents and antisolvents used in perovskite film fabrication

| Type | Short name | Full name | Density [g/ml][77] | Boiling point[77] [°C] | Dipole moment [77] [D] | $\delta_D$ | $\delta_P$ [MPa$^{1/2}$][78,79] | $\delta_H$ |
|---|---|---|---|---|---|---|---|---|
| S | DMF | Dimethyl Formamide | 0.95 | 153 | 3.86 | 17.4 | 13.7 | 11.3 |
| S | ACN | Acetonitrile | 0.79 | 82 | 3.92 | 15.3 | 18.0 | 6.1 |
| S | NMP | N-methyl-2-pyrrolidone | 1.03 | 202 | 4.1 | 18.0 | 12.3 | 7.2 |
| S | DMSO | Dimethyl Sulfoxide | 1.1 | 189 | 3.96 | 18.4 | 16.4 | 10.2 |
| S | GBL | Gamma-Butyrolactone | 1.13 | 204 | 4.27 | 18.0 | 16.6 | 7.4 |
| S | PC | Propylene Carbonate | 1.20 | 242 | 4.9 | 20.0 | 18.0 | 4.1 |
| S | 2ME | 2-Methoxy-ethanol | 0.96 | 124 | 2.04 | 16 | 8.2 | 15.0 |
| S | THF | Tetrahydrofuran | 0.89 | 65 | 1.63 | 16.8 | 5.7 | 8.0 |
| S | DMAc | Dimethyl-Acetamide | 0.94 | 165 | 3.72 | 16.8 | 11.5 | 9.4 |
| S | DMPU | 1,3-dimethyl-3,4,5,6-tetrahydro-2(1H)-pyrimidinone | 1.06 | 247 | 4.17 | 17.8* | 5.1* | 9.3* |
| S | DMI | 1,3-dimethyl-2-Imidazolidinone | 1.05 | 225 | 4.05 | - | - | - |
| S | 2P | 2-pyrrolidone | 1.12 | 245 | 3.7 | 18.2 | 12.0 | 9.0 |
| S | H2O | Water | 0.99 | 100 | 1.85 | 15.5 | 16.0 | 42.3 |
| A/S | EtOH | Ethanol | 0.79 | 78 | 1.69 | 15.8 | 8.8 | 19.4 |
| A/S | IPA | 2-Propanol | 0.79 | 83 | 1.66 | 15.8 | 6.1 | 16.4 |
| A/S | BuOH | 1-Butanol | 0.81 | 118 | 1.66 | 16.0 | 5.7 | 15.8 |
| A/S | IBA | Isobutyl Alcohol | 0.80 | 108 | 1.66 | 15.1 | 5.7 | 15.9 |
| A/S | t-BuOH | Tert-Butanol | 0.78 | 82 | 1.31 | 15.2 | 5.1 | 14.7 |
| A/S | EA | Ethyl Acetate | 0.9 | 77 | 1.88 | 15.8 | 5.3 | 7.2 |
| A/S | CF | Chloroform | 1.49 | 61 | 1.15 | 17.8 | 3.1 | 5.7 |
| A/S | CB | Chlorobenzene | 1.11 | 131 | 1.69 | 19.0 | 4.3 | 2.0 |
| A/S | BA | n-Butyl Acetate | 0.88 | 126 | 1.87 | 15.8 | 3.7 | 6.3 |
| A/S | oDCB | O-Dichloro-benzene | 1.3 | 180 | 2.5 | 19.2 | 6.3 | 3.3 |
| A/S | Ani | Anisole | 1.0 | 154 | 2.3 | 17.8 | 4.4 | 6.9 |
| A/S | TFT | Trifluorotoluene | 1.19 | 103 | 2.86 | 17.5 | 8.8 | 0 |
| A/S | DCM | Dichloromethane | 1.33 | 39.6 | 1.6 | 17.0 | 7.3 | 7.1 |
| A/S | DEE | Diethyl Ether | 0.71 | 35 | 1.15 | 14.5 | 2.9 | 4.6 |
| A/S | mXyl | m(eta)-Xylene | 0.86 | 139 | 0.33-0.37 | 17.4 | 1.0 | 3.1 |
| A/S | pXyl | p(ara)-Xylene | 0.86 | 138 | 0 | 17.4 | 1.0 | 3.1 |
| A/S | oXyl | o(rtho)-Xylene | 0.89 | 144 | 0.64 | 17.8 | 1.0 | 3.1 |
| A/S | Tol | Toluene | 0.87 | 111 | 0.36 | 18 | 1.4 | 2.0 |
| A/S | Mesit | Mesitylene | 0.86 | 164.7 | 0.047 | 18 | 0.6 | 0.6 |
| A/S | DMC | dimethyl carbonate | 1.07 | 90 | 0.93 | 15.5 | 8.6* | 9.7 |
| A/S | DEC | diethyl carbonate | 0.97 | 125 | 0.91 | 15.1* | 6.3* | 3.5* |
| A/S | -- | Hexane | 0.66 | 69 | 0.09 | 14.9 | 0 | 0 |
| A/S | -- | Octane | 0.70 | 126 | 0 | 15.5 | 0 | 0 |
| A/S | -- | Cyclohexane | 0.78 | 81 | 1.16 | 16.8 | 0 | 0.2 |
| A/S | 2MA | 2-Methylanisole | 0.98 | 171 | 1.0 | - | - | - |
| A/S | MeTHF | 2-methyltetrahydrofuran | 0.85 | 80 | 1.38 | 16.9 | 5.0 | 4.3 |
| A/S | LM | (R)-(+)-limonene | 0.84 | 176 | 0.2-0.7 | 17.2 | 1.8 | 4.3 |
| A/S | DPE | Diphenyl ether | 1.07 | 258 | 1.17 | 19.4 | 3.4 | 4.0 |

## Hansen Solubility Parameter Model – History and Underlying Principles

### Solubility parameters, the conventional approach

The term *solubility parameter* (δ) was first used by Hildebrand and Scott, with further contribution of Scatchard's early work for its development.[80,81] Shortly after its introduction, the Hildebrand solubility parameter, became the standard for solubility parameter and was defined as the square root of the cohesive energy density ( □ = $(E/V_m)^{1/2}$, E is the energy of vaporization and $V_m$ is the molar volume of the pure solvent). These cohesive energies arise from interactions of a given solvent molecule with another one of its own kind.[78]

The Hildebrand's solubility parameters are widely implemented as the solubility standard in different fields in diverse industries to aid in the selection of solvents for coating preparations, help in the prediction of compatibility of polymers, their chemical resistance, and permeation rates, and even to characterize the surfaces of pigments, fibres and fillers.[82–84] Liquids with similar solubility parameters will be miscible, and polymers will dissolve in solvents whose solubility parameters are not too different from their own. The basic principle has been referred as *"like dissolves like"*. However, more recently, this concept has been modified to *"like seeks like,"* as various surface characterizations have also been made, and surfaces do not (usually) dissolve. In short, solubility parameters help put numbers into a simple qualitative idea which results is a quantity for predicting solubility relations.

**Thermodynamic basis of solubility parameters**. When considering the miscibility of two fluids, which can be melts of organic materials, or the solubility of a polymer in a solvent, thermodynamic requirements have to be fulfilled. The free energy of mixing ($\Delta G^M$), which is defined as the difference of the enthalpy and entropy changes, must be zero or negative. The contribution of temperature is introduced by factoring it with the entropy change. The first theory for the calculation of the entropic and enthalpic contributions was published by Flory and Huggins.[85,86] For a binary mixture made of two solvents, the free energy of mixing is defined by Equation (1),

$$\Delta G^M = RT \cdot (\phi_1 ln\phi_1 + \phi_2 ln\phi_2 + \chi_{12}\phi_1\phi_2) \quad (1)$$

with the consideration of the universal gas constant *R*, the temperature *T*, the volume fractions φ and the Flory-Huggins interaction parameter $\chi_{12}$. Nevertheless, the determination of the parameter $\chi_{12}$ is enormously complicated and is not possible to obtain from typical material characteristics. The $\chi_{12}$ parameters for specific material combinations, can be rarely used in a generalised way while the experimental effort to obtain them is high, so that alternative approaches to calculate the enthalpic interaction gained focus. Among those approaches, the most accepted one for decades was introduced in 1936 and further developed in 1950 by Hildebrand and Scott.[80,81] They defined the free energy of mixing depending on the cohesive energy *E*, the volumes of the components and the volume of the mixture $V_{12}$. Furthermore, the cohesive energy was defined as the total energy necessary to break all intermolecular physical bonds of a substance under isothermal and isobaric conditions. Although it is not referring to any specific molecular interaction, Hildebrandt's approach allowed the estimation of the $\chi_{12}$ parameter, making it possible to determine solubility parameters of many substances. Although widely used for decades, the implementation of Hildebrand's solubility parameter has its limitations. For instance, larger molecular species do not follow the prediction. Even for substances having identical solubility parameters, the heat of mixing as proposed has been shown to be incorrect. In general, its use is limited to regular solutions, not accounting for connection between molecules, such as those that polar and hydrogen bonding interactions would require. To overcome these limitations, multicomponent solubility parameters have been proposed. Among them the Hansen Solubility Parameters have been regarded as a more expanded approach, as it considers atomic and molecular interactions.

**Hansen Solubility Parameter Space**

The Hansen Solubility Parameters theory firstly introduced by Dr. Charles M. Hansen in 1967 and more comprehensively described in 2007, further develops Hildebrand and Scott's solubility parameters theory, taking into account the contributions to the energy balance of molecular interactions coming from (atomic) dispersion forces ($\delta_D$), (molecular) permanent dipole-permanent dipole forces ($\delta_P$), and (molecular) hydrogen bonding ($\delta_H$).[47,78] Subsequently, considering the early definition of $\delta$ as the result of partial contributions of molecular interaction as proposed by Hansen, it leads to a more comprehensive description of the cohesive energy and solubility, which is the corresponding sum of the dispersive, polar, and hydrogen bonding components (cohesion energy Equation (2) and solubility Equation (3) after dividing (2) by the molar volume),

$$E = E_D + E_P + E_H \quad (2)$$
$$\delta^2 = \delta_D^2 + \delta_P^2 + \delta_H^2 \quad (3)$$

From Equation (3) and the respective partial solubility parameters of two substances (indices 1 and 2), the interpretation of similarity, or affinity, between them becomes possible by calculating their solubility parameter "distance" $R_a$ following Equation (4),

$$R_a^2 = 4(\delta_{D1} - \delta_{D2})^2 + (\delta_{P1} - \delta_{P2})^2 + (\delta_{H1} - \delta_{H2})^2 \quad (4)$$

where $R_a$ can be seen as the distance between two points in a three-dimensional space. For this, the factor "4" in Equation (4) resulting in a doubling of the dispersion parameter axis leads to a spherical appearance of the solubility volume instead of an ellipsoidal shape without that correction. The HSPs are temperature dependent. In particular, $\delta_H$ is temperature sensitive, as more and more hydrogen bonds brake progressively at higher temperatures. Nevertheless, the accuracy of predictions is reliable at room temperature and correlations of phenomena at higher temperatures have been found to be sufficiently explainable and described with solubility parameters established at 25 °C.[78]

The HSP theory is also based on the idea that "*like seeks like*", consequently if any two substances have very similar HSPs, a high interaction between them is expected. Thus, through the HSP model it is possible to predict if two substances are compatible, miscible or exhibit high affinity with each other. Particularly in the case of polymer solubility, the accuracy of HSP has been debated as for polymer dissolution not only temperature plays a role in solubility but also the solvent's molecular size, permeation and diffusion phenomena. In particular, the molecular size and chemical structure influences diffusion kinetics and thus, the HSPs can be insufficient to predict solubility of polymers. Nevertheless, unlike solid-fluid mixtures, for single organic molecules including liquid-liquid mixtures the influence of the molecule's/molecules' structure on the solubility is less significant as there are fewer significant diffusion phenomena involved.

In the case of various substances placed simultaneously in the Hansen space, an interaction sphere can be described using the partial solubility parameters ($\delta_D$, $\delta_P$, $\delta_H$) of the substance of reference as the coordinates of the centre, while the distance between molecules ($R_a$) serves as the radius, whereas any other substances can be plotted as points in the Hansen space, this is illustrated in Figure 2.

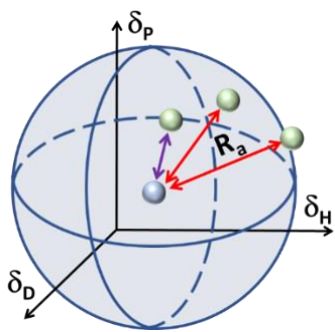

**Figure 2.** Schematic of the Hansen interaction space for a given chemical compound (blue dot) and its interaction distance (Ra) in relation to other chemical compounds (green dots).

The Hansen space thus provides a visual representation of the solubility between two fluids. For instance, the farther the compounds are to the reference compound's interaction sphere (i.e. larger $R_a$ in Figure 2), the less likely it is that one substance is soluble in the other. Inversely, the closer they are in the Hansen space, the more probable it is that the substances in comparison are soluble (e.g. the solvent pointed with purple line). Traditionally, $R_0$ is known as the 'interaction radius' and is the magnitude of interaction of a molecule within the Hansen space. Typically, $R_0$ of solutes is obtained experimentally using known 'good' and 'bad' solvents. Reports on the calculation of $R_0$ are available whereas the method for its determination specifically for solvents is arguably not fully settled.[87] Yet, according to some authors the accurate determination of $R_0$ for most compounds remains empirical and requires data-fitting.[88,89] In a more general way, $R_0$ can be defined as the maximum reach of interaction of a molecule in the Hansen space starting from its HSP centre. This way, two compounds will interact if their spheres touch or overlap. In the explicit case of two solvents in a liquid state, solvents closer to each other in the Hansen Space are most likely to be miscible and thus form a homogeneous mix when combined. It is worth mentioning that the HSPs for most common compounds have been already experimentally obtained and applied following the methodology described by Hansen and Abbott.[78,90] Nowadays they are found in literature, databases, datasheets and new procedures for numerical estimation of the parameters have been proposed,[91] making the implementation of HSPs a simple method to use for early predictions of solubility and other applications. Currently, HSPs are successfully used in the painting and coating industry, as aid for substitution with less hazardous formulations in various other types of products such as cleaners, printing inks and adhesives. More recently, their usefulness has been reported even in improving printable energetics.[92]

Perovskite inks contain high concentrations of ionic precursors and the HSP theory does not account for ionic interactions of salts nor molecular complexing in these relatively highly concentrated inks. In the search to include these considerations other researchers report the introduction of further parameters and their combination with HSP. Some attempts have been reported towards solvent selection models based on donor number (DN), dielectric permittivity or dipole moment incorporated in new models as an additional solubility parameter[87,93]. Particularly, Lei et al. have adapted DN in terms of cohesive energy ($\delta_{DN}$)[87] to quantify the coordinate cohesion between $PbI_2$ and donor solvents which in combination with HSPs aids in the understanding of interactions between Lewis acid solutes (e.g., halide salts) and solvents via alternative solubility models. In their work, they determined a range of values for $\delta_{DN}$, $\delta_P$ and $\delta_H$, at which higher quality perovskites can be produced. However, as remarked by Di Girolamo et al., in contrast to HSPs for which there's a vast library of HSPs (for over 10, 000 molecules) plus a well-defined method to determine them, DN has limited availability in literature which makes the use DN as a generalized parameter to find suitable solvents unpractical.[94] Moreover, DN is based on Lewis basicity and in a molecular picture, a non-charged solvent molecule can only be a Lewis base if it provides electron density via lone-pairs and exhibits an intrinsic dipole moment. These two molecular properties however are already covered by two of the three HSP, namely $\delta_P$ and $\delta_H$, describing polar interactions and hydrogen bonding (that would occur via free electron pairs), respectively. Mathematically speaking – DN is most likely a function of $\delta_P$ and $\delta_H$, while the exact mathematical connection among the three variables remains so far unknown. Nevertheless, all these efforts also show the increasing need to establish reliable and generalized quantitative methods to determine better suitable solvents for halide perovskite precursors, solvent/antisolvents and their mixtures.

Perovskite systems are multi-component systems and interactions among solvent(s), solute and antisolvents can be rather complex. HSP-theory, however, also allows here an approach to understand interactions among these components beyond a simple binary system.

To judge interactions among different components of a mixed system, comparing the position in HSP of the individual components with respect to the corresponding interaction sphere (of radius $R_0$) is crucial. To understand stronger or preferential interactions among various components of a perovskite mixture the use of the Relative Energy Differences (RED) in the Hansen space, given by $R_a/R_o$ was suggested by some authors[87]. RED allows to rank interactions among various substances. Strong interactions are characterised by a RED<1 while weak or no interactions would be resembled by RED≥1.

Conversely, affinity of different mixtures can still be obtained by comparing their distances in the Hansen space. Moreover, as the creators of the HSP theory propose, it is possible to elucidate the HSP of mixtures if the HPS of the other component and their concentration in the mix are known.[95] The HSP of the mix is the volume-weighted average of the HSP of the individual components. For instance, Figure 3 shows a hypothetical 'Solute' which is poorly soluble by either solvent A or solvent B, represented by their position outside of the corresponding interactions sphere. Once solvents A and B are mixed, for example in a 1:1 volume ratio, new HSPs can be calculated that resembles the HSP of the solvent mixture. If this new HSP-point lies within the interaction sphere of that solute, it will dissolve. Homologically, this methodology can be implemented for other solvent combinations and for solvent/antisolvent/additives systems. Since the new set of HSP of the mixture is the volume weighted average of the individual components, dilute solutions or small amounts of additives in for instance an antisolvents can be treated as the plain original

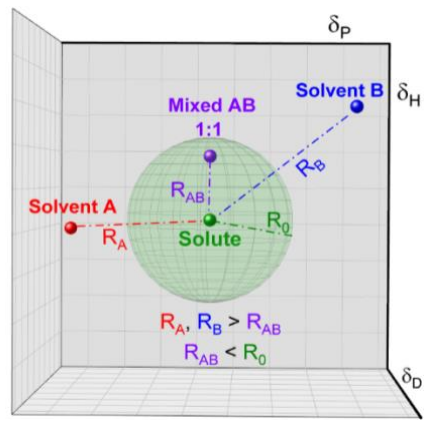

solution.

**Figure 3.** Representation of the distances of a mixed compound in relation to a fixed solute.

## HSP theory applied to MHP film formation

### Solvent-antisolvent interactions

As explained above, interactions between compounds can be inferred using HSPs from their distance (Ra) and this can be applied to improve MHP film formation. In the case of single solvent system thus a small Ra between solvent and antisolvent is desirable plus low interactions to the perovskite precursors. Then the removal of host solvent from the wet film will trigger supersaturation efficiently leaving the perovskite composition unmodified.

To date, most efforts applying HSP theory in MHP have focused on quantifying the solubility for commonly used perovskite precursor salts in order to enhance interactions between the precursor solution and the halide salts.[75,76,96,97] Yet, until recently no attention has been placed on employing it to improve the solvent-engineering method towards controlling the film formation process, by, for example, a selective removal of solvents in precursor solution mixtures. It is in these cases were employing HSPs theory can become a great instrument to quantitatively select antisolvents and/or solvents that are beneficial for more efficient solvent engineering methodology.

It has been shown by us that photovoltaic devices fabricated from highly oriented MHP structures are superior in stability and performance as compared to those with random polycrystalline structures.[67] Recently, we reported that the formation of highly oriented triple cation perovskite films can be triggered by the use of alcoholic antisolvents. For example, Figure 4a shows the comparison of the microstructures of triple cation perovskite thin films fabricated using either TFT or IPA as antisolvents. The SEM images show that the grains in the IPA-based films are terminated by flat surfaces (marked in pink), suggesting a preferred crystalline orientation. On the other hand, in the case of TFT, a more random orientation of the grains is observed, since various different grain terminations can be seen (marked in maroon). In situ GIWAXS measurements (Figure 4b) allow to monitor the crystallisation process in real time and reveal the formation of a short-lived intermediate directly after the application of IPA as an antisolvent, which impacts on the subsequent crystallisation of the perovskite layer. This intermediate species is not present for the TFT case. Consequently, GIWAXS patterns collected on the completed perovskite layers (Figure 3c) show a significantly larger degree of preferred orientation in the case of films processed from IPA. Additional experiments revealed that the intermediate phase consists of a FAI-$PbI_{2-x}$·DMSO complex, that forms as a consequence of the preferential removal of DMF from DMF:DMSO host solvent mixtures during the solvent-engineering process when using alcoholic antisolvents.[67]

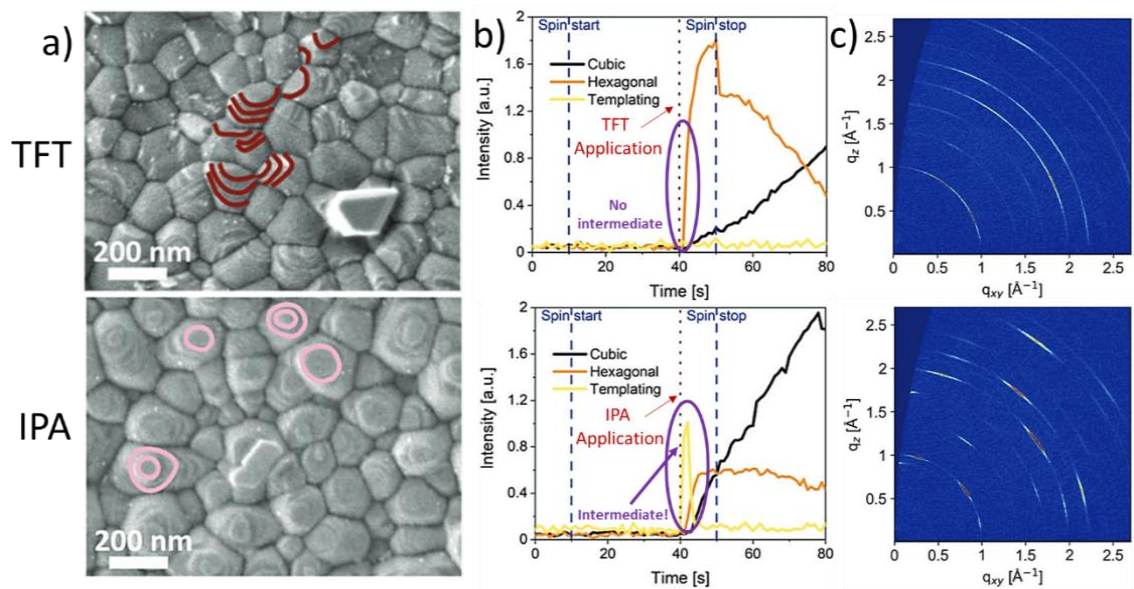

**Figure 4.** Differences in crystallization and templating intermediate formation between TFT (top) and alcohols (IPA as example in the bottom). a) Scanning electron microscope (SEM) images collected via the secondary electron detector of triple cation perovskites, b) evolution of the templating species, the hexagonal and cubic phases and c) final grazing-incidence wide-angle-ray scattering obtained post-annealing. [Adapted with permission from Telschow et al.,[67] Adv. Sci. 2206325 (2023). Copyright 2023 Author(s), licenced under a Creative Commons Attribution Licence].

These observations are supported by considering the Hansen solubility parameters of the solvents involved in the film fabrication process, as increased hydrogen bonding between the alcohols and the polar DMF host solvent promote a stronger interaction of DMF with the alcoholic antisolvents than DMSO. Contrarily, when considering a non-alcoholic solvent such as TFT, which has similar $\delta_D$ and $\delta_P$ to DMF and DMSO and no contribution in $\delta_H$, the result is an equally good extraction of both host solvents, inhibiting the creation of the highly oriented intermediate phase. Additionally, in our work, the Ra of DMF ($R_{a\_DMF}$) and DMSO ($R_{a\_DMSO}$) against different solvents was compared by calculating the ratio $R_{a\_DMF}/R_{a\_DMSO}$. HSP theory reveals a correlation between the above-mentioned improved features on the MHP film related to the intermediate species and the magnitude of the $R_{a\_DMF}/R_{a\_DMSO}$ ratio.

Alcohols as antisolvents present low and similar ratios (~0.74) which is significantly larger than for the non-alcohols (~0,96). This shows that a selective removal of one solvent by means of solubility analysis of the solvent-antisolvent mixture can have a significant impact towards improving the performance of metal halide perovskite-based cells when implementing HSP theory in solvent engineering fabrication.

**Improving metal halide perovskite-based cells via HSP**

Following the aforementioned potential of interaction-oriented solvent-antisolvent strategies to control the formation and orientation of polycrystalline perovskite films, applying the Hansen Solubility Parameters could also be a promising approach towards tailoring perovskite films with high homogeneity and improved crystal microstructures. In this context, some concepts must be firstly defined, including: "What is the solvent-antisolvent desired proximity in the Hansen space? How to determine the extent of solvent-antisolvent interaction spheres that is beneficial for solvent removal? Is it possible to design adequate multi solvent-antisolvents mixtures?" and other open questions that remain unanswered. Recent works have addressed how implementing HSP estimation could also help substituting undesirable solvents with different alternatives. By substituting solvents, it could become possible to reduce the used amount of certain undesired solvents and instead select more favourable ones within a desired proximity in the Hansen space. Among the first studies employing HSP theory for MHP was the study by Gardner et al. in which the authors identified non-toxic solvents for the fabrication of $MAPbI_3$ films.[98] In this direction, Wang et al. used HSPs to determine the distance in Hansen space between commonly used $MAPbI_3$ precursor solvents and the assumed parameters of the lead iodide solute.[99] They found that it is possible to find alternative solvents (such as 2-MP) besides the traditional well known DMF and DMF/DMSO mixtures used for deposition of perovskite structures.

Additionally, from the alternative solvents obtained from HSPs, other selection criteria can be included. For instance, a co-solvent combination favouring lower boiling point solvents, which would help reduce the energy needed to dry the perovskite layer, or more effective substitution of hazardous solvents for greener ones. More specifically, for improving perovskite crystallization in solvent-antisolvent-halide salt mixtures, to achieve the highly selective removal of one solvent in solvent mixtures, the desired proximity in Hansen space must be close enough to ensure that there's higher affinity between the antisolvent and the solvent intended to be removed first than that intended to be removed later (i.e., $R_a$ of solvent to remove < $R_a$ solvent to keep). Wang et al. also highlighted the relevance of knowing the chemical affinity of solvents and antisolvents as the purity of the intermediate phase after the solvent removal depends on their chemical affinity with the antisolvents.[97] The authors observed a clear difference in the formed intermediate phase structures depending on the antisolvent used where they failed to subtract unwanted solvent. They attributed this occurrence to the relatively similar polarities of chlorobenzene (CB) and anisole (Ani) making them miscible with both DMF and DMSO respectively. As suggested by Telschow et al., the similar solubility of antisolvents reduces the effectivity of selective removal of the host solvent accelerating the saturation of the precursor solution and preventing formation of the intermediate phase that promotes a higher degree of crystallinity.[67] These recent reports highlight the complexity of solvent-antisolvent interactions involved in the process for removing undesired solvents and retaining the antisolvents that aid promoting higher efficiency MHP-based PV devices. Thus, a more well-defined methodology must be proposed in order to fully take advantage of the HSP theory towards improving PV devices.

**Strategic tailoring of solvent-antisolvent mixtures**

To this day, despite the evidently high impact of solvent-antisolvent solubility via the solvent-engineering method in determining the quality of the perovskite film and the subsequent performance of the PV devices made from them, a quantitative approach in solvent engineering towards finding the optimal solvent-antisolvent mixtures has not been proposed. Given how much the solubility interactions in solvent-antisolvent mixtures directly influence film formation, crystallization, crystal orientation, homogeneity, and defects, new approaches are needed in solvent-engineering methodology to further improve the stability while retaining high performance of metal halide perovskite photovoltaic devices. Since HSP theory considers the polar, dispersive and hydrogen bond interactions, employing the Hansen theory in solvent-engineering methods would bring particular advantages for improved MHP films.

For instance, it would enable the possibility to strategically select better solvents for perovskite precursor solutions besides the most typical ones. Consequently, it provides the possibility to apply the same methodology in any kind of photosensitive materials like lead-free perovskites, organic-inorganic halide perovskites, multi-tandem cells, tin-based or mixed lead-tin-based perovskites and other hybrid materials. It could also aid in the selection of antisolvents that allow full removal of solvents and facilitate intermediate phase formation. Implementing HSP theory would also enable designing co-solvent/co-antisolvent mixtures that promotes higher quality microstructured films. All while the preliminary assessments can be done solvent-free through computer modelling.

To exemplify the potential of HSP theory for designing solvent-antisolvent mixtures, the HSPs of the most common solvents and antisolvents employed in the fabrication of metal halide perovskite films were plotted (Figure 5a) using data from Table 1. In addition, three spheres were inferred from the visualization, grouping the species as follows: host solvents (red), aprotic antisolvents (blue) and antisolvent alcohols (green), Figure 5b. For purposes of this example, the boundaries of the spheres were assumed based on the observed proximity of clustered compounds.

To achieve high quality perovskite layers from intermediate phase formation, a pre-condition is selecting an adequate solvent-antisolvent combination. Thus, thinking for instance of a DMSO/DMF/halide salts precursor solution where the subsequent selective removal of DMF is known to favour the intermediate phase formation, it would require an antisolvent with a closer Hansen proximity to DMF than to DMSO. Based on this requirement, it is possible to visualize from Figure 5b that potential favourable antisolvents are for instance antisolvents inside the alcohols sphere. This observation is in full agreement with the study reported by Telschow et al. where DMF was preferentially removed when 2-propanol (IPA), 1-butanol (BuOH) or isobutyl alcohol (IBA) were used as antisolvents.[67]

An additional advantage of implementing HSPs in metal halide perovskite film fabrication, is being able to identify greener solvents that can be used for perovskite precursor solutions and more energy efficient processing routes. This is particularly important since the overall mass of solvent used for perovskite film formation and therefore their environmental impact is orders of magnitude higher than that of the solid materials.[100,101] Furthermore, drying and evaporation of solvents via thermal annealing are the main contributor to the energetic fabrication cost of solution-processed MHP-devices.

Some reports have proven the potential of using HSP theory to lower the toxicity of MHP films, not only by substituting hazardous chemicals by ones with low toxicity, but also by predicting the compatibility of low toxicity solvents with salt precursors towards a greener fabrication of metal halide perovskite-based PV devices.[54,75,96]

In this context the Hansen solubility model can be highly effective in finding suitable, non-hazardous solvents to solubilize lead salts for the fabrication of lead halide perovskite films by providing insight into the interaction between the salts and the solvents employed, as suggested by Dooling et al.[96]

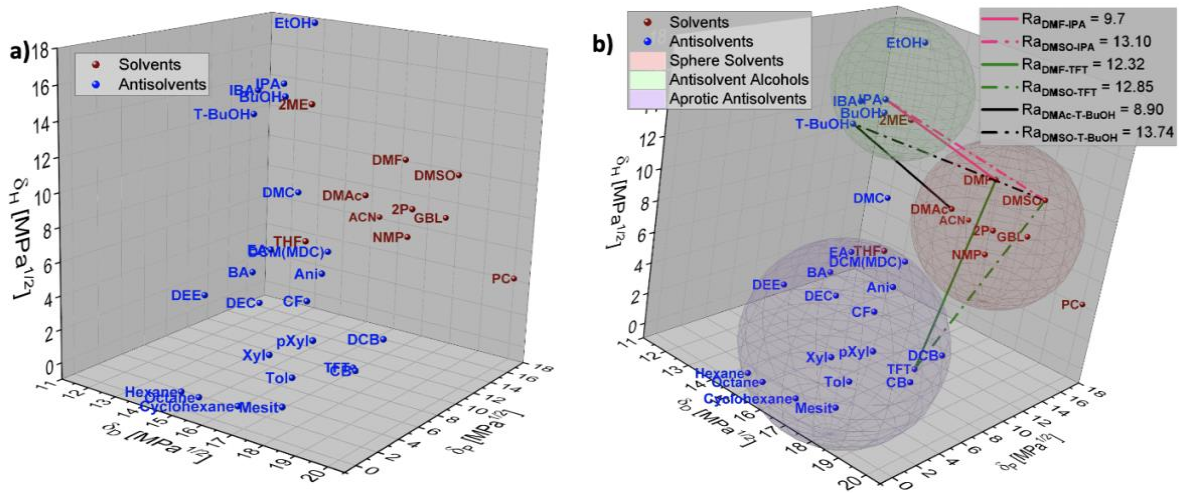

**Figure 5.** Hansen solubility space including a) solvents and antisolvent, b) spheres enclosing the most common solvents (red), antisolvents (blue) and antisolvent alcohols (green) used in perovskite film fabrication with Ra distances plotted for specific solvents and antisolvents combinations.

The substitution of hazardous solvents for greener ones would aid avoiding unintended consequences from adopting alternative solvent systems Visually inspecting the plotted Hansen space in Figure 5b, candidates to substitute DMF can be elucidated, such as DMAc. DMAc is a known solvent for halide salts,[55] similar in properties to DMF, as shown in Table 1, yet it presents lower toxicity than DMF, making it a possible alternative to substitute DMF in perovskite layer fabrication. Moreover, inside the alcohol antisolvents sphere in Figure 4b, *t*-BuOH appears to be in the 'desired proximity' as defined previously. Following equation (4) the distance of IPA to DMF and DMSO is 9.70 and 13.10 respectively, while the distance of *t*-BuOH to DMAc and DMSO is 8.90 and 13.74, confirming that this antisolvent is closer to DMAc and farther to DMSO than IPA is to DMF and DMSO. This is indicating that the combination DMAc:DMSO+*t*-BuOH could work better to promote the formation of the halide perovskite intermediate phase while simultaneously reducing the toxicity of the solution mixture.

Following these principles and the current state of the art in solvent-engineering methods, implementing solubility parameters models such as the HSP model and visualizing the 3D Hansen space with available data for solubility predictions can become a powerful and resource-saving tool for designing solvent-antisolvent combinations towards improved PV devices, while as well having potential to aid in reducing the environmental impacts of their fabrication. Nevertheless, it is worth noticing that there are challenges to consider. One being the non-existence of HSP data for some chemicals (yet obtainable experimentally as described in literature)[78,102,103] and the accuracy for more complex compounds.

This method, yet intricate, shows the tailoring capabilities that implementing solubility models brings to solvent-engineering methodology towards improved perovskites. Moreover, as will be discussed in the following section, the miscibility of the antisolvent and host solvents should also be accounted for.

Overall, selecting the optimal solvent-antisolvent combination for perovskite solar cell fabrication requires careful consideration of its solubility, toxicity, boiling point and other properties. The use of alternative solvents with lower toxicity and cost has the potential to improve the stability, scalability and commercial viability of halide perovskite solar cells while upholding their high efficiency. Although, further research is needed to explore the use of new solvents and antisolvent combinations and their impact on the performance of perovskite solar cells, the use of HSP models for their prediction is highly promising.

**Antisolvents with very low solvent interactions**

As discussed above, solvent-antisolvent interactions can have a dramatic impact on the crystallization mechanisms and dynamics and consequently affect the microstructure, grain orientation and composition of the resulting perovskite film. Very weak interactions among the solvent(s) and antisolvent(s) may lead to poor solvent-antisolvent miscibility. Antisolvents that are poorly miscible with the commonly employed polar host solvents like DMF, DMSO, GBL and NMP, are typically aprotic, non-polar solvents. The intermolecular interactions of such "poorly-miscible" antisolvents are dominated by dispersive forces, a very low intrinsic dipole moment and their molecular structures are typically characterized by the absence of functional groups to form strong hydrogen bonds. Typical representatives for such non-polar solvents used as antisolvents for MHP are toluene,[39] xylene,[41,104] mesitylene,[72] dialkyl ethers,[105,106] hexane,[107] or other hydrocarbons.[108,109] In the Hansen parameter space these poorly-miscible solvents cluster in opposing sections of the HSP space when compared to the polar host solvents DMF, DMSO, GBL and NMP. Consequently, the weak interactions of the non-polar antisolvents with the host solvents are characterized by large distances in Hansen space $R_a$. Only the group of alkyl ethers are somewhat closer to the typical polar solvents for perovskites, suggesting a better miscibility with the host solvents. The determination of the overall miscibility of a solvent-antisolvent system can be complicated by the fact, that often mixtures of host solvents (e.g. DMF:DMSO) are used, with varying degrees of miscibility with the antisolvent. Diethyl ether (DEE) for example, is still miscible with DMF at room temperature, but is basically immiscible with DMSO. Since the HSP do not allow to define a threshold distance for which miscibility is still observed, the simplest way to assess the miscibility of the solvents employed in the fabrication of perovskites is to mix them in the appropriate ratios experimentally and observe the outcome. For non-miscible solvents a phase boundary will form or the solvent mixture will turn turbid, indicating phase separation on much smaller length scales. While the occurrence of phase boundaries indicates non-miscibility, the exact composition of each phase cannot simply be derived from such experiments, and small volumes of host solvents might be soluble in larger amounts of antisolvent. Furthermore, the temperature of the precursor solution and the antisolvent can strongly affect miscibility, offering the potential to control their interaction.

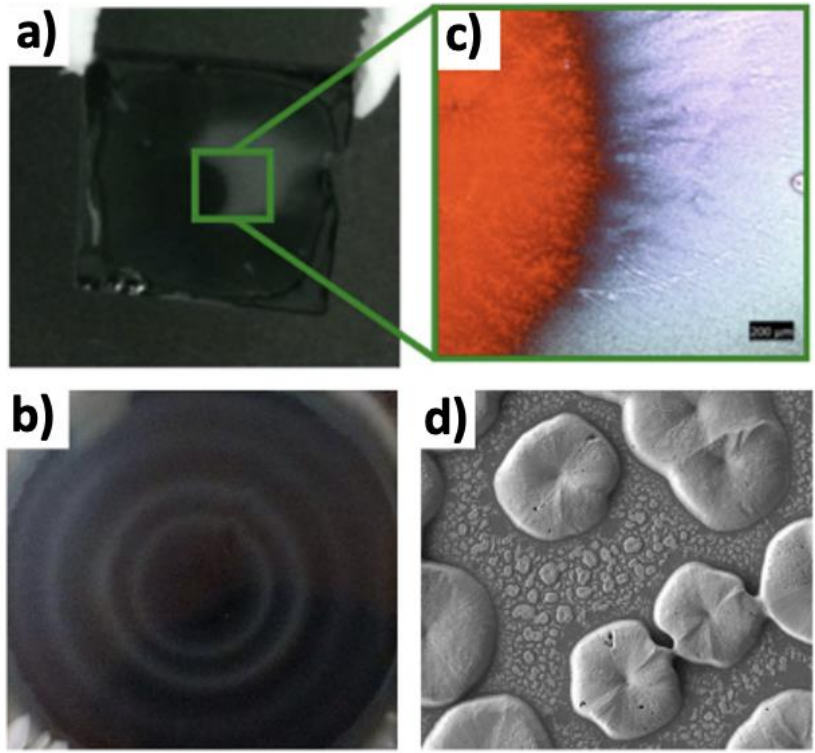

**Figure 6.** Inhomogeneities and microstructural flaws of perovskite films after using a non-miscible antisolvent. a) film with toluene as antisolvent b) radial film variations using xylene c) optical microscopy images of an area of film in a) highlighting the lack of a perovskite film outside the inner substrate center d) SEM of a perovskite film using mesitylene showing discontinuous film formation. [Panel 6b adapted with permission from Manion et al.[99] Copyright 2020 American Chemical Society. Panels 6a, c, f adapted with permission from Taylor et al.,[67] Nat. Commun. 9, 1878 (2021). Copyright 2021 Author(s), licenced under a Creative Commons Attribution Licence].

Importantly, even antisolvents that are poorly-miscible with the host solvent(s) can be employed in the fabrication of high-quality perovskite layers and for the extraction of host-solvents, if the application parameters are chosen correctly. Due to the low miscibility of non-polar antisolvents with the host solvents, their overall removal and the initialization of the nucleation and solidification process during the antisolvent application are very different than for other antisolvents. Unlike the more polar antisolvents, the fraction of the host-solvent that the poorly-miscible, non-polar antisolvents can remove from the wet film during their application, is very limited. On the other hand, the solubility of the lead, caesium and organic halide precursors in such antisolvents is also typically very low. These properties together require the antisolvent to be applied slowly, ideally with a continuous flow as was recently demonstrated for triple cation perovskites[72] and MA-free perovskites.[57] Alternatively, larger volumes of such non-miscible antisolvents need to be chosen to achieve supersaturation.[110]

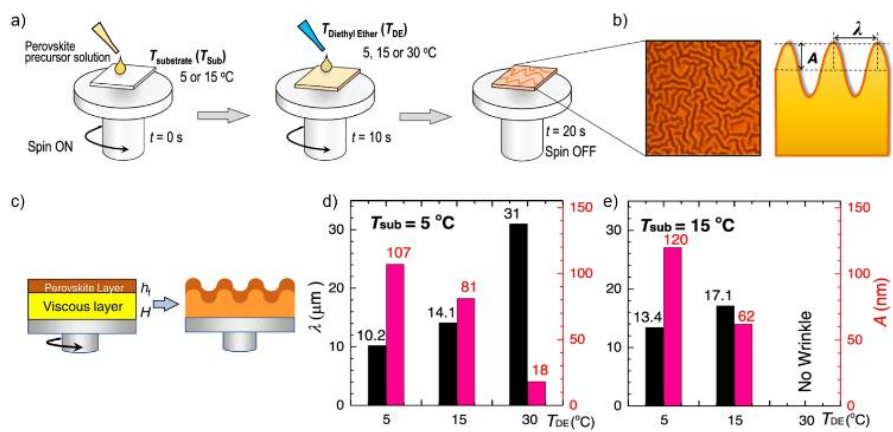

**Figure 7**. a) Schematic application of diethyl ether (DE) at various temperatures of substrate ($T_{Sub}$) and antisolvent ($T_{DE}$) b) optical microscopy and schematic cross section of wrinkled perovskite film c) intermediate formation of the elastic capping layer on the viscoelastic bottom solution upon application of a non-miscible antisolvent d) and e) properties of the wrinkled films shown in depending on $T_{Sub}$ and $T_{De}$. [Adapted with permission from Kim et al.,[100] Nat. Commun. 12, 1554 (2021). Copyright 2021 Author(s), licenced under a Creative Commons Attribution Licence].

Both strategies can ensure successful removal of the host solvent and initiate the crystallization process, while leaving the final perovskite composition unchanged. Deviating from this processing window for non-miscible antisolvents, for example by applying the antisolvent too fast or in too small quantities, will result in a highly defective microstructure with large inhomogeneities (Figure 6).[72,110] For spin-coating this often results in non-compact, inhomogeneous films with radial strongly varying film quality (Figure 6a and b). Often the central area still resembles (after thermal annealing) a perovskite film of high quality with low roughness (Figure 6c). However, areas off the centre are particularly rough, discontinuous and exhibit an island type growth (Figure 6d). It remains unclear, however, if these flaws in microstructure are a result of a local phase segregation of the solvent mixture or of the incomplete and inhomogeneous conversion of the perovskite precursor solution into the desired intermediates and pre-perovskite crystallites. In a recent study by Kim et al. the miscibility of diethyl ether as antisolvent and DMSO as solvent was correlated with the appearance of wrinkles in the resulting perovskite, a microstructural feature that is often observed in perovskite layers (see Figure 7).[111] Wrinkles are periodic height variations of the perovskite films with distinct wavelength and amplitude (Figure 7b) The authors could show in their combined experimental and theoretical study that the use of diethyl ether as antisolvent leads to the formation of a solid capping layer on the wet viscous precursor solution (Figure 7c). The resulting bilayer structure and the differences in the mechanical and elastic properties of the solid and viscous layers result in the wrinkling of the film. The characteristic length scale and amplitude of the film wrinkling could be controlled by the temperature dependent miscibility of diethyl ether in DMSO (Figure 7d,e).

At low temperatures the two solvents are non-miscible, resulting in a strong vertical segregation and consequently strong wrinkling, while for an antisolvent temperature around 30°C the miscibility of DMSO and diethyl ether is higher, resulting in a much more homogenous extraction and no wrinkles. A high degree of miscibility of the antisolvent with the host solvent simplifies to a certain extent the fabrication of high-quality perovskite layers, yet the use of non-miscible solvents can still result in high quality films as long as the application parameters are considered carefully. Since the miscibility is strongly temperature dependent, the influence of temperature on the resultant film microstructure and quality will be substantially higher in the case of non-polar antisolvents. Nevertheless, the use of immiscible antisolvents may introduce microstructural variations on large length scales and needs to be considered carefully for large-area film formation.

**Role of precursor solubility in antisolvents**

While in general, antisolvents are meant to not dissolve the precursors of the perovskite layers, in certain cases, some of the precursors might be soluble or partly soluble in them. This means that during the application of the antisolvent, a partial removal of these precursors might occur, thus irreparably altering the stoichiometry of the perovskite layer and oftentimes, adversely impacting its microstructure due to the formation of pinholes or voids (Figure 8a). For example, MAI is highly soluble in alcoholic antisolvents and to some degree even in ethyl acetate or other chlorinated antisolvents (Figure 7b).[72] This suggests that in addition to considering the solvent-antisolvent interactions, it is important to consider the antisolvent-precursor interactions as well. In particular, we have shown that the speed of the antisolvent application, consequently the duration for interaction, is a crucial parameter that can compensate for the unintended removal of perovskite precursor during deposition. A very rapid antisolvent application enables the formation of high-quality perovskite films even by antisolvents that are capable of dissolving the perovskite precursors, due to the limitation in the interaction time. Consequently, solar cells fabricated using a fast antisolvent application rate resulted in reasonable performance as high as 20%, while those fabricated slowly lead to a far lower performance and higher degree of performance variation (Figure 8c). These considerations would need to be introduced into the HSP model in order to enable the selection of alternative antisolvents whilst accounting for precursor solubility and its impact on antisolvent application parameters.

**Figure 8.** (a) surface and cross-sectional scanning electron microscopy images of perovskite

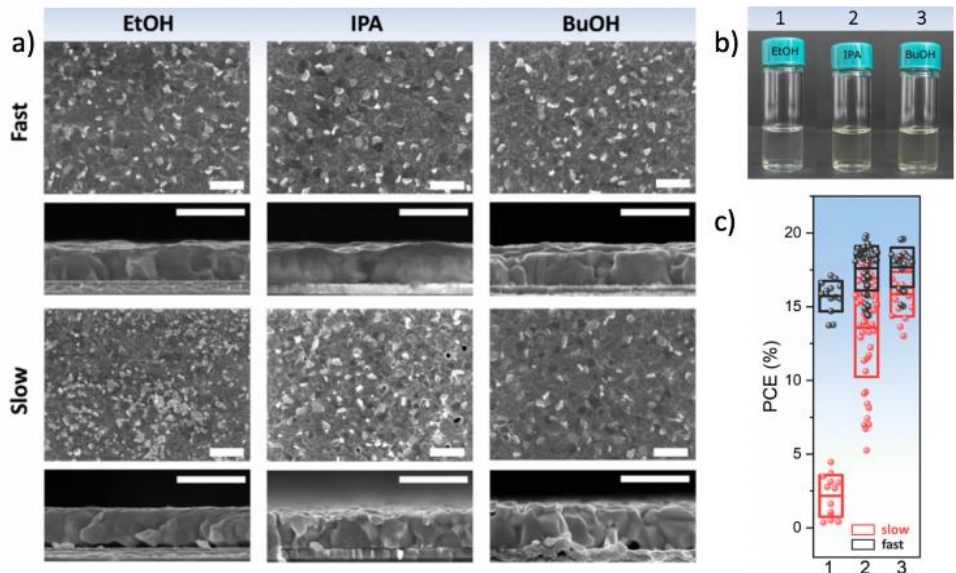

films formed from Type I antisolvents (EtOH, IPA, and BuOH). Scale bar is 1 μm. (b) solubility of MAI in a solution of DMF:DMSO:antisolvent, demonstrating that MAI is soluble in Type I antisolvents, (c) the power conversion efficiency of solar cells resulting from a fast or slow application of Type I antisolvents. [Adapted with permission from Taylor et al.,[72] Nat. Commun. 9, 1878 (2021). Copyright 2021 Author(s), licenced under a Creative Commons Attribution Licence].

**Antisolvent categorisation and impact on processing conditions**

Based on the outlined above, both solvent-antisolvent and precursor-antisolvent interactions must be considered as factors that can impact on the film formation process, namely:

(1) the solubility of the antisolvent with the host solvents;
(2) the miscibility of the antisolvent with the host solvents;
(3) the solubility of the perovskite precursors in the antisolvent.

These three factors were empirically found to influence the quality of the perovskite films, and consequently the performance of perovskite solar cells.[57,67,72] Importantly, a study by Taylor et al. demonstrated that these factors have significant implications for the choice of optimal processing conditions of the layers.[72] There, the 14 investigated antisolvents were categorised into Type I, Type II and Type III based on the optimal antisolvent application rate that led to high solar cell performance(Figure 9a).

Specifically, Type I antisolvents were found to lead to high quality films and high-performance devices only when the antisolvent was applied fast, while a slow application led to many microstructural flaws and excess of $PbI_2$. Type II antisolvents led to high quality films and good photovoltaic performance regardless of the speed of their application. Finally, Type III antisolvents were best applied slowly.

Examining these empirical observations through the perspective of the Hansen space allows to develop a predictive suggestion for the categorisation of antisolvents, thus not only eliminating the need for the time and effort consuming solar cell fabrication, but also providing guidelines towards the optimal application procedure of antisolvents. Figure 8b visualises the 14 antisolvents from Taylor et al.[72] in the HSP space, demonstrating that by considering the HSPs their categorisation is simplified and no longer requires extensive experimental characterisation. Type I antisolvents, namely alcohols are well separated from the other antisolvents due to the high contribution of the hydrogen bonding $\delta_H$ component. Type III antisolvents are characterised by low hydrogen bonding and polar forces ($\delta_P$), are generally immiscible with DMSO:DMF consequently placing them predominantly at one quarter of the ($\delta_D$, $\delta_P$) plane of the Hansen space. Finally, the versatile group of the Type II antisolvents includes contributions from all 3 HSPs.

For all three types of antisolvents, the depicted sphere was selected to enclose as many solvents from each antisolvent type in order to provide the smallest experimentally confirmed boundary for antisolvent categorisation. Using these spheres, other antisolvents – beyond those that were included in the original study – can now be categorised without extensive experimental characterisation if they fall within their boundaries. Moreover, once categorised, the rate of antisolvent application can be directly chosen to match the requirements of each antisolvent type, thus leading to high efficiency devices with minimal optimisation.

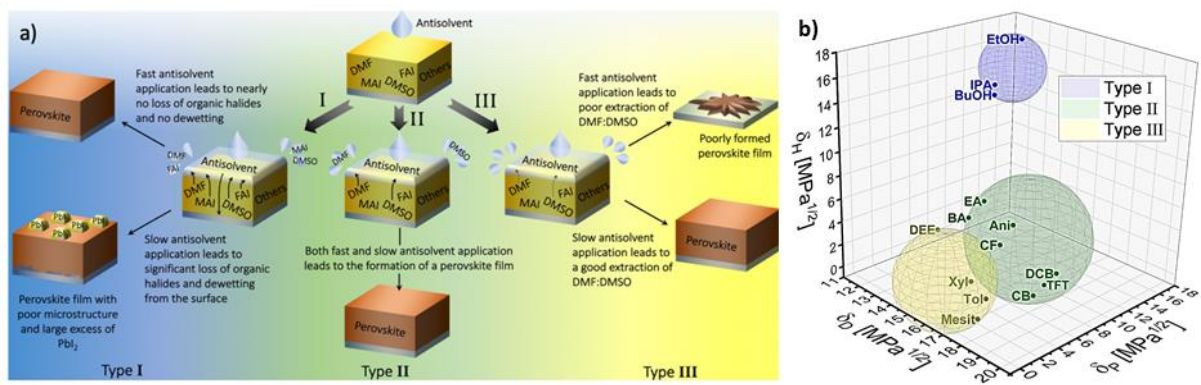

**Figure 9.** (a) Illustration of the processes involved in the film formation of perovskite films made by different types of antisolvents when the antisolvent is applied fast and slow. (b) Visualization in the Hansen space of Type I, Type II and Type III antisolvents. Boundaries of each sphere was selected to enclose as many solvents from each corresponding antisolvent type. [Panel 8a reprinted with permission from Taylor et al., Nat. Commun. 9, 1878 (2021). Copyright 2021 Author(s), licenced under a Creative Commons Attribution Licence].

We note that the boundaries of the spheres currently demonstrate the lower limit for each category and it is possible that with additional experimental data, their boundaries can be extended. Importantly, for antisolvents falling outside the sphere boundaries, it is still possible to predict their categorisation based on their proximity to the spheres, but experimental validation in this case would be necessary.

**Future research directions and impact on application of perovskites in large-scale optoelectronic devices**

Despite the significant insights into the role of antisolvents in triggering the crystallisation of perovskite layers and enabling control over the resultant microstructure and crystalline structure, many open questions remain unanswered. For example, much of the existing body of literature focuses on the study of film formation in lead-based MHP, while the role of antisolvents in the crystallisation of tin-based or mixed lead-tin-based perovskites is far less investigated.[112–114] This could be related to the fact that crystallisation in the latter occurs on much faster timescales, which complicates its study.[115,116] On the other hand, the increasing availability of in-situ methods such as in-situ optical characterisation[117] by absorption and photoluminescence spectroscopy and in-situ structural characterisation by GIWAXS[60] opens the path to investigate crystallisation processes also in these perovskites. While it is likely that the importance of solvent-antisolvent miscibility and solubility will impact crystallisation also in case of tin-based perovskites, additional factors may come into play and would need to be identified.

Another fundamental question is related to the potential efficacy of antisolvent mixture engineering as a method for controlling crystallisation and passivation of perovskite thin films. For example, while the use of mixtures of host solvents (e.g. DMF:DMSO) is highly common, the use of antisolvent mixtures remains largely unexplored.[118] Based on the solubility of perovskite precursors in the various antisolvents, and their Hansen solubility parameters, it is possible to utilize antisolvent mixtures that specifically target a particular film formation process. For instance, Telschow et al demonstrated that using a mixture of 1:1 of TFT and IPA as an antisolvent enables the formation of highly oriented films (as is observed by using solely IPA), yet with a reduced risk of altering the stoichiometry of the perovskite layer, which may occur due to the high solubility of organic halides in IPA.[67] This example illustrates that the beneficial properties of several solvents can be combined to achieve high quality crystalline films with the desired stoichiometry and orientation. Importantly, the use of antisolvent mixtures may also enable the widening of the processability windows by relaxing the restrictions related to the application rates of different antisolvents.

Another advantage of utilizing antisolvent mixtures is the ability to introduce molecular passivation agents directly in the antisolvent step. In such a case, the passivating molecule might not be directly soluble in the desired antisolvent, yet it can be introduced by utilising an antisolvent mixture. For example, Degani et al used a 9:1 mixture of CB and IPA in order to introduce a series of 2-phenylethylammonium iodide salts (X-PEAI) as additives for passivation.[119] These salts are soluble in IPA, but are not soluble in CB and the use of CB:IPA mixtures enabled their introduction without altering the crystallisation dynamics of the perovskite layer. Detailed spectroscopic characterisation revealed that the use of the antisolvent application step for the introduction of passivation agents leads solely to the modification of the top surface of the perovskite layer,[120] suggesting that it cannot substitute the need for bulk additives that are commonly added to the perovskite precursor solution.[121,122] The passivation of the top surface, however, was highly efficient, with more than a doubling of the photoluminescence quantum yield for films formed by PEAI containing antisolvent mixtures. This example and other similar works illustrate the benefits of utilising the antisolvent step to not only guide crystallisation, but simultaneously passivate or functionalise the surface of the perovskite layer.[123,124]

Beyond the fundamental understanding of crystallisation and film formation processes of solvent engineered films and the engineering of antisolvent mixtures and compositions, it is important to consider the applicability of the approach for large-scale production of perovskite optoelectronics. While the use of antisolvents is well established for small area devices, its application on large-scale devices leads to significant engineering and environmental challenges. The former is related to the fact that spin-coating is not a suitable method for large area fabrication of MHPs, so the processing and the application of antisolvents would need to be adapted to the new deposition technology. Moreover, for large area processing, very large volumes of antisolvents would need to be applied, which may not only trigger the formation of inhomogeneities,[125,126] but also impact the crystallisation dynamics through an increased antisolvent vapour above the substrate. A promising technique in this case can be the spraying of the antisolvent, which proved promising both on small and large areas of samples.[59,127] However, significantly more research is required in order to fully adapt the solvent-engineering method for large area fabrication.

Beyond the technological challenges that need to be addressed, as highlighted before, the high toxicity of many of the antisolvents may pose significant health risks and environmental hazards, thus negatively impacting the life-cycle analysis of perovskite optoelectronic devices.[101,128] This motivated the search for greener chemicals that can be utilized for perovskite film formation.[97,129] However, typically, such antisolvents lead to a slightly inferior film quality and a reduced device efficiency. Significantly more research into the processing of perovskites using green antisolvents is required in order to ensure that the safety and environmental concerns are resolved, without sacrificing device performance.

**Conclusion**

In a nutshell the solvents, antisolvents or additive selection guidelines can be summarised as follows:

- Once all the compounds that will be part of the solution system are defined, the first step is to identify the requirements in terms of the desired outcome (e.g., selective removal, precipitation, solvent substitution, etc.).
- Next, with their HSPs visualize and estimate the distances in the Hansen space and determine additional parameters that are relevant to the process (e.g., boiling point).

- Subsequently, narrow down the selection by categorizing the solvents and antisolvents (and other additives in the case of more complex systems) according to the common features and miscibility, this is 'the selection criteria'.
- Finally, by visualizing the substances in the solubility model it becomes possible to recognise potential candidates that comply with the previously defined selection criteria and the identified features wanted like selective removal of a solvent using a specific antisolvent).

Although based in complex solubility theory, it is the simplicity and potential that the application of solubility models as the proposed here that brings a clearer path towards improved and feature-tailored perovskites fabrication methods as well as improved categorisation of solvents and antisolvents. Many fundamental and applied research questions related to MHP film formation by solvent engineering are yet to be resolved. The visualisation of antisolvents in the Hansen space, as well as the additional insights and guidelines discussed above can be used to direct the search for new, alternative antisolvents, that would be environmentally benign, yet highly effective in achieving the desired MHP film quality. Similarly, this approach can also be used for the investigation of antisolvent mixtures that would enable a controlled engineering of the structure and microstructure of the perovskite layers.

## Authors Contributions

Conceptualization, J.R.B.Q., F.P. and Y.V.; Investigation, O.T. and J.R.B.Q. Resources, Y.V. and F.P; Supervision, F.P. and Y.V.; Writing - original draft, J.R.B.Q.; Writing - review & editing, O.T., F.P. and Y.V. All authors have read and agreed to the published version of the manuscript.

## Conflicts of interest

The authors declare no conflicts of interest.

## Acknowledgements

The authors thank the Deutsche Forschungsgemeinschaft (DFG) for funding the "PERFECT PVs" project (Grant No. 424216076) via the Special Priority Program 2196. Y. V. has received funding from the European Research Council (ERC) under the European Union's Horizon 2020 research and innovation programme (ERC grant agreement number 714067, ENERGYMAPS).